\documentclass[%
onecolumn
 reprint,
superscriptaddress,
 amsmath,amssymb,
 aps,
]{revtex4-2}

\usepackage{graphicx}
\usepackage{epstopdf}
\usepackage{dcolumn}
\usepackage{bm}
\usepackage{physics}
\usepackage{enumitem}
\usepackage{xcolor}
\usepackage{hyperref}

\begin{document}

\title{Nonlocality of the energy density of a spontaneously emitted single-photon from a Hydrogen atom}

\author{M. Federico}
\affiliation{Telecom Paris, Institut Polytechnique de Paris, 19 Place Marguerite Perey, 91120 Palaiseau, France}
\affiliation{Laboratoire Interdisciplinaire Carnot de Bourgogne - ICB, Universit\'e de Bourgogne, CNRS UMR 6303, BP 47870, 21078 Dijon, France}
\author{H.R. Jauslin}
\email{jauslin@u-bourgogne.fr}
\affiliation{Laboratoire Interdisciplinaire Carnot de Bourgogne - ICB, Universit\'e de Bourgogne, CNRS UMR 6303, BP 47870, 21078 Dijon, France}
\date{\today}%

\begin{abstract}
We analyze through the expectation value of the energy density the spatial nonlocality of single photons emitted by the spontaneous decay of a Hydrogen atom. By using a minimal coupling between the quantized electromagnetic field and the atom, we compute the state of the photon under the assumption that only a single-photon is produced. The calculations are thus performed in the subspace of single-photon states which is essentially equivalent to the rotating wave approximation. We obtain a characterization of the spatial decay of the energy density. We compute the asymptotic limit of large distances from the atom at each given time, and find an algebraic behavior of $1/r^6$. This result confirms that the energy density of single-photon states is nonlocal and the algebraic decay is far from the maximal quasiexponential localization predicted by the theory.
\end{abstract}

\maketitle

\section{Introduction}

The description of the spontaneous decay of a Hydrogen atom is a standard calculation that can be done using either the Weisskopf-Wigner theory \cite{Weisskopf1930,Spohn2009,Novotny2012} or time-dependent perturbation theory, e.g., with the Fermi golden rule \cite{Novotny2012}. However, these approaches in general consider a dipolar coupling between the field and  the atom, taken as a two-level system, which prevents from computing the state of the emitted photons since there is a frequency Lamb shift given by a diverging intergral. The treatment of this divergence could be done by a renormalization procedure, as suggested in \cite{Derezinski2002}. This problem was discussed, e.g., in \cite{Moses1973,Moses1975,Moses1975a,Grimm1974,Grimm1975,Seke1988} and a way out was found by considering a minimal coupling instead of the dipole approximation, which provides an intrinsic ultraviolet cutoff function that avoids the need of a renormalization. In this article we analyze the spatial properties of the single-photon states that would be produced in this process. In view of the intrinsic spatial nonlocality of single-photons, as discussed, e.g., in \cite{Knight1961,Hegerfeldt1974,Hegerfeldt1994,Hegerfeldt1997,Hegerfeldt1998,BialynickiBirula1998,BialynickiBirula2009, Federico2023a}, we would like to characterize this nonlocality from the perspective of spontaneous emission. To do so, we will consider the approach that uses the energy density to probe the nonlocality of photons, as it was originally done in \cite{BialynickiBirula1998,BialynickiBirula2009} and recently extended to all single-photon states in \cite{Federico2023a}. In this article it was shown that for any single-photon state, the expectation value of the energy density in a small volume is different from zero anywhere in space. 
This result leads to several natural questions if we consider the photons produced by some concrete source: (i) If one assumes that what is produced are single-photon states, what would be their spatial localization properties? (ii) How does the mean value of the energy density depend on the distance between the source and the detector? 
We analyze these questions with the example of the spontaneous emission of a Hydrogen atom prepared in its first excited state.
We start with the complete  Hamiltonian of the atom interacting with the quantized electromagnetic field. The hypothesis that only a single-photon is produced is implemented by projecting the model to the single-photon subspace. Furthermore we keep only the fundamental and the first excited state  of the atom, corresponding to the Lyman-$\alpha$ transition. The time evolution in the Schr\"odinger picture can be represented in terms of  a system of Weisskopf-Wigner equations that can be solved exactly in terms of some integrals \cite{Spohn2009}. This provides the spatio-temporal description of the photon state, which we analyze at each fixed time as a function of the distance from the source. 
The expectation value of the electromagnetic energy density is non-zero in any finite volume at arbitrary distances  $r$ from the source, as it was proven for all single-photon states in \cite{Federico2023a}.
The asymptotic analysis for large distances leads to the result that at each fixed time the energy density is proportional to $1/r^6$. 

We conclude that the spontaneous emission cannot produce a strictly single-photon state since it would contradict causality \cite{Berman2004,Dolce2006}: At any fixed time $t$ there would be a nonzero probability of detecting the photon with an instrument located at an arbitrary distance $r$ from the source, in particular one could take $r> ct$, where $c$ is the speed of light \cite{Gulla2021,Ryen2022,Gulla2023}.
We remark that we made the analysis for the situation where the atom is in the first excited state at $t=0$, but a similar argument can be made if at $t=0$ the atom is in a linear combination of the ground state and the first excited state. This could correspond to the state prepared with a $\pi$-pulse of finite duration.

Several works have already addressed related questions in the context of spontaneous emission \cite{Hegerfeldt1994,Biswas1990,Buchholz1994,Keller2000,Chan2002,Fedorov2005}. In particular, causality properties have been analyzed in \cite{Debierre2015a,Debierre2015,Debierre2016,Debierre2018}, where different regimes, in the near- mid- and far-field regions were described in terms of wave functions constructed by Glauber's extraction principle from a first order correlation function. This wave function is identical, up to a global constant, with the Bia{\l}ynicki-Birula representation of single-photon states \cite{BialynickiBirula1996}, which is in turn equivalent to the representation of the states that we will use in this work (Landau-Peierls) \cite{Federico2023}. However, we note that the formulas (25) in Ref. \cite{Debierre2016} do not provide the asymptotic behavior for large distances $r=\abs*{\vec x}$, since they still contain integrals over $k$ involving functions of $\abs*{\vec x}$ that will modify the asymptotic form for large $\abs*{\vec x}$. Moreover, we stress that our spatial analysis will not be performed for the state of the field but for its energy density expectation value. This is a major difference since it has been shown (see, e.g., \cite{Federico2023a}) that in some particular situations, the state function can be spatially localized even though the corresponding energy density features a nonlocality. The nonlocality property is thus not a property of the states alone but of the joint representation of the states and the corresponding local observables.

Causality properties were also analyzed in \cite{Compagno1990} where it was shown using the electric energy density observable that the nonresonant terms are needed for the spontaneous emission to be causal. This result is in agreement with the characterization we perform in this work for the resonant part, for which we also provide an expilcit asymptotic behavior. Such an asymptotic analysis, performed through perturbation theory calculations, was also done in \cite{Compagno1995} but for the dressed vacuum only, i.e., atom in its ground state and zero photons. 

Finally, as it was noted by Shirokov \cite{Shirokov1978}, since the eigenfunctions of the Hydrogen atom decrease exponentially in the radial variable, they are nonzero everywhere. Therefore the emision of photons cannot be expected to be local, but should have at least an exponential tail. This effect has been described as a blurring of the causal light cone. However, this is an exponential decay which is a relatively small effect compared to the nonlocality properties described below.

To analyze the spatial properties of the photon states, we will use throughout this work a position space formulation of the quantum theory of the electromagnetic field called the Landau-Peierls representation (see Appendix \ref{LP and helicity} for its definition). Any other position space representation could in principle be used and the result is independent of this choice \cite{Federico2023}.

The structure and the main results of the article are as follows. In Section \ref{WW model} we formulate the Weisskopf-Wigner model with the $\vec{\hat A}\cdot \vec{\hat p}$ interaction and we use the vector spherical harmonics and helicity basis for single-photon states, as proposed in
\cite{Moses1973,Moses1975,Moses1975a,Grimm1974,Grimm1975,Seke1988}. This basis has the advantage that it leads easily to exact selection rules for the interaction with Hydrogen atoms. 
In Section \ref{int solution} we write the Weisskopf-Wigner solution in terms of integrals. This solution is exact under the condition that the spectrum of the associated Friedrichs operator is absolutely continuous, which is expected in the small coupling regime. Some of these integrals can be calculated explicitly if one makes a resonant approximation which has essentially the same effect as the Weisskopf-Wigner approximation, but without extending an integral to non-physical negative frequencies.
With the coupled atom-electromagnetic field state, we calculate in Section \ref{expec ener} the expectation value of the energy density at an arbitrary position in $\mathbb{R}^3$.
We then analyze its dependence on the radial distance, in particular for the far-field, which characterizes the global extension of the nonlocality. We obtain the result that the expectation value of the energy density is proportional to $1/r^6$ for large $r$, i.e., it has an algebraic decay, which is far from the theoretically maximal quasiexponential decay predicted in \cite{BialynickiBirula1998}.

\section{Weisskopf-Wigner model}\label{WW model}

We consider a non-relativistic Hydrogen atom represented by a two-level system corresponding to the Lyman-$\alpha$ transition, i.e., with ground state $\ket{n_g=1;j_g=0;m_g=0}\equiv\ket{g}$ and excited state $\ket{n_e=2;j_e=1;m_e=0,\pm1}\equiv\ket{e}$. This transition is relevant since it can be prepared experimentally by applying, e.g., a resonant $\pi$-pulse on the ground state with a linearly polarized laser light for $m_e=0$ and a circularly polarized laser light for $m_e=\pm1$. The wavefunctions associated to the ground and excited states are, respectively, 
\begin{equation}\label{hyd eigfun}
\ket{g}=\varphi_g(r,\vartheta,\varphi)=\frac{1}{\sqrt{\pi r_\text{B}^3}}e^{-\frac{r}{r_{\text{B}}}},\quad\quad\quad\ket{e}=\varphi_e(r,\vartheta,\varphi)=\beta_{m_e}(\vartheta,\varphi)\frac{r}{r_{\text{B}}}e^{-\frac{r}{2r_{\text{B}}}},
\end{equation}
where $r_{\text{B}}=\hbar/(\alpha mc)$ is the Bohr radius expressed, here, in terms of the fine structure constant $\alpha$, the electron mass $m$, and the speed of light $c$. The functions $\beta_{m_e}(\vartheta,\varphi)$ depend on the choice that is made for $m_e=\{0,\pm1\}$ and read
\begin{equation}
\beta_0(\vartheta,\varphi)=\frac{1}{4\sqrt{2\pi r_{\text{B}}^3}}\cos\vartheta,\quad\quad\quad
\beta_{\pm1}(\vartheta,\varphi)=\mp\frac{1}{8\sqrt{\pi r_{\text{B}}^3}}\sin\vartheta \ e^{\pm i\varphi}.
\end{equation}
The free Hamiltonian of the two-level atom can be expressed as
\begin{equation}
\hat H_{\text{at}}=E_g\ket{g}\bra{g}+E_e\ket{e}\bra{e},
\end{equation}
where $E_g$ and $E_e$ are the energies of the ground and excited states, respectively.
To describe the spontaneous emission, the two-level Hydrogen atom is interacting with the quantized electromagnetic field which we write using the basis $\ket{k,J,M,\lambda}\equiv\vec\psi_{k,J,M}^{(\lambda)}$ of helicity vector spherical harmonics (see Appendix \ref{helicity functions} for more details about these functions)
\begin{equation}\label{free elm}
\hat H_{\text{em}}=\int_0^\infty dk\sum_{J,M}\sum_{\lambda=\pm}\hbar\omega_k\hat B^\dag_{\vec\psi_{k,J,M}^{(\lambda)}}\hat B_{\vec\psi_{k,J,M}^{(\lambda)}}.
\end{equation}
We remark, here, that the creation-annihilation operators $\big(\hat B^\dag_{\vec\psi_{k,J,M}^{(\lambda)}}, \hat B_{\vec\psi_{k,J,M}^{(\lambda)}}\big)$ are defined on functions in the Landau-Peierls representation \cite{Landau1930,Federico2023} (see Appendix \ref{LP and helicity}), and fulfill the following commutation relation $\big[\hat B_{\vec\psi_{k,J,M}^{(\lambda)}}, \hat B_{\vec\psi_{k',J',M'}^{(\lambda')}}^\dag\big]=\delta(k-k')\delta_{J,J'}\delta_{M,M'}\delta_{\lambda,\lambda'}$.
For the interaction part, and following \cite{Moses1973}, we take an operator of the form
\begin{equation}
\hat H_{\text{int}}=-\frac{e}{m}\vec{\hat p}\cdot\vec{\hat A}(\vec x),
\end{equation}
where $\vec{\hat p}=-i\hbar\nabla$ is the momentum operator of the electron and $\vec{\hat A}(\vec x)$ is the vector potential operator in the Coulomb gauge. We have neglected the $\hat A^2$ term coming from the minimal coupling since it has been shown by Moses \cite{Moses1973} (see also \cite{Grimm1974,Grimm1975}) that taking an interaction of the form $\vec{\hat p}\cdot\vec{\hat A}$ is enough to have a consistent model allowing to compute the photon state. Moreover, the typical order of magnitude for the effects related to the $\hat A^2$ term have been estimated to be much smaller than those from the $\vec{\hat p}\cdot\vec{\hat A}$ term \cite{Mandel1995}.
The resulting total Hamiltonian is finally
\begin{subequations}
\begin{align}
\hat H&=\hat H_{\text{at}}+\hat H_{\text{em}}+\hat H_{\text{int}}\\
&=E_g\ket{g}\bra{g}+E_e\ket{e}\bra{e}+\int_0^\infty dk\sum_{J,M}\sum_{\lambda=\pm}\hbar\omega_k\hat B^\dag_{\vec\psi_{k,J,M}^{(\lambda)}}\hat B_{\vec\psi_{k,J,M}^{(\lambda)}}-\frac{e}{m}\vec{\hat p}\cdot\vec{\hat A}(\vec x).
\end{align}
\end{subequations}

In the following, since we are interested in the characterization of single photons, we project the system into the subspace generated by the vacuum and the single-photon states, which corresponds in the tensor product Hilbert space to the subspace generated by 
$$
\{ \ket{e;\varnothing},\ket*{g;\vec\psi_{k,J,M}^{(\lambda)}} \},
$$
i.e., the atom in its excited state and zero photons and the atom in its ground state and one photon. This approximation is essentially equivalent to the rotating wave approximation that is performed in general for the dipole coupling. Within this subspace, the interaction coefficient takes the form \cite{Moses1973}
\begin{subequations}
\label{coupling}
\begin{align}
\rho(k,J,M,\lambda)&=-\frac{e}{m}\bra*{g;\vec\psi_{k,J,M}^{(\lambda)}}\vec{\hat p}\cdot\vec{\hat A}(\vec x)\ket*{e;\varnothing}\\
&=\left(\frac{2}{3}\right)^{7/2}\sqrt{\frac{\alpha^5}{\pi}}mc^2\frac{k/K}{\sqrt{k}}\frac{\delta_{J,1}\delta_{M,m_e}}{\left[1+\left(\frac{k}{K}\right)^2\right]^2}\\
&\equiv\rho(k)\delta_{J,1}\delta_{M,m_e},
\end{align}
\end{subequations}
where $K=3/(2r_{\text{B}})$. 
Here, we have chosen a slightly different convention for the global phases of the eigenfunctions $\vec\psi_{k,J,M}^{(\lambda)}$ than the one of \cite{Moses1973}. Our convention has the advantage that $\rho$ is real and independent of the helicity $\lambda$.

\section{Solution in terms of integrals}\label{int solution}

Starting with a general state in the subspace of interest
\begin{equation}
\ket{\Psi}=c_0\ket{e;\varnothing}+\int_0^\infty dk\sum_{J,M,\lambda}c_{k,J,M,\lambda}\ket*{g;\vec\psi_{k,J,M}^{(\lambda)}},
\end{equation}
we want to solve the Schrödinger equation 
\begin{equation}
i\hbar\frac{\partial}{\partial t}\ket{\Psi}=\hat H\ket{\Psi},
\end{equation}
with the initial condition $c_0(t=0)=1$ and $c_{k,J,M,\lambda}(t=0)=0$, for any $k,J,M$ and $\lambda$. Expressed for the nonzero coefficients, we arrive at the set of equations
\begin{subequations}
\begin{align}
i\hbar\dot c_0(t)&=E_ec_0(t)+\int_0^\infty dk\sum_{\lambda=\pm}\rho(k)c_{k,\lambda}(t),\\
i\hbar\dot c_{k,\lambda}(t)&=\rho(k)c_0(t)+\left(E_g+\hbar\omega_k\right)c_{k,\lambda}(t),
\end{align}
\end{subequations}
where we have introduced the shorter notation $c_{k,J=1,M=m_e,\lambda}\equiv c_{k,\lambda}$. Moreover, since $\rho(k)$ is $\lambda$-independent, the last equation is also $\lambda$-independent and thus $c_{k,+}(t)=c_{k,-}(t)$ for all $t$ and one can define $d_k(t)=\sqrt{2}c_{k,\lambda}(t)$, to rewrite the set of dynamical equations as
\begin{subequations}
\label{set}
\begin{align}
i\dot c_0(t)&=\omega_{\text{a}}c_0(t)+\int_0^\infty dk\frac{\sqrt{2}}{\hbar}\rho(k)d_k(t),\\
i\dot d_k(t)&=\frac{\sqrt{2}}{\hbar}\rho(k)c_0(t)+\omega_kd_k(t),
\end{align}
\end{subequations}
where we have set the energy of the ground state to $E_g=0$, and introduced the Bohr frequency of the two-level system $\omega_{\text{a}}=(E_e-E_g)/\hbar$. Expressed in terms of $\omega_k=ck$, equations (\ref{set}) can be written in a matrix form
\begin{equation}
\label{Friedrich model}
i\begin{bmatrix}\dot c_0\\\dot D_{\omega_k}\end{bmatrix}=\begin{bmatrix}\omega_{\text{a}}&\bra{\tilde\rho}\ket{\cdot}\\\tilde\rho&\omega_k\end{bmatrix}\begin{bmatrix}c_0\\D_{\omega_k}\end{bmatrix},
\end{equation}
where $D_{\omega_k}=d_{\omega_k}/\sqrt{c}$ and
\begin{equation}
\tilde\rho(\omega_k)=\sqrt{\frac{2}{c\hbar^2}}\rho(\omega_k).
\end{equation}
This formulation is known as the Friedrichs-Lee model \cite{Friedrichs1948,Lee1954} and the exact solution in terms of integrals can be formulated as \cite{Spohn2009}
\begin{subequations}
\label{Friedrich solution}
\begin{align}
c_0(t)&=\int_0^\infty d\omega\ g(\omega)e^{-i\omega t},\\
D_{\omega_k}(t)&=-ie^{-i\omega_kt}\tilde\rho(\omega_k)\int_0^t dt'\ e^{i\omega_kt'}c_0(t'),
\end{align}
\end{subequations}
with
\begin{equation}\label{g factor}
g(\omega)=\frac{1}{2\pi}\frac{\Gamma(\omega)}{(\omega-\omega_e-\Delta(\omega))^2+\frac{\Gamma(\omega)^2}{4}},
\end{equation}
and
\begin{equation}\label{decay+freq shift}
\Gamma(\omega)=2\pi\abs*{\tilde \rho(\omega)}^2,\quad\quad\quad
\Delta(\omega)=\text{pv}\int_0^\infty d\omega_k\ \frac{\abs*{\tilde \rho(\omega_k)}^2}{\omega-\omega_k},
\end{equation}
where $\text{pv}$ is the principal value.
In the weak coupling regime, $g(\omega)$ has a high peak centered at $\omega=\omega_{\text{a}}$ and one can thus make the following approximation \cite{Spohn2009}
\begin{subequations}
\begin{align}
g(\omega)\simeq g_{\text{w}}(\omega)&=\frac{1}{2\pi}\frac{\Gamma(\omega_{\text{a}})}{(\omega-\omega_{\text{a}}-\Delta(\omega_{\text{a}}))^2+\frac{\Gamma(\omega_{\text{a}})^2}{4}}\\
&=\frac{1}{2\pi}\frac{\Gamma_{\text{a}}}{(\omega-\omega_{\text{a}}-\Delta_{\text{a}})^2+\frac{\Gamma_{\text{a}}^2}{4}},
\end{align}
\end{subequations}
where $\Gamma_{\text{a}}=\Gamma(\omega_{\text{a}})$ and $\Delta_{\text{a}}=\Delta(\omega_{\text{a}})$. This approximation allows to express $c_0(t)$ as
\begin{equation}
c_0(t)\simeq\int_0^\infty d\omega\ g_{\text{w}}(\omega)e^{-i\omega t}=e^{-\frac{\Gamma_{\text{a}}}{2}t}e^{-i(\omega_{\text{a}}+\Delta_{\text{a}})t},
\end{equation}
yielding a result that is essentially equivalent to the Weisskopf-Wigner approximation. From this, one can further compute 
\begin{equation}\label{Dk}
D_{\omega_k}(t)\simeq -ie^{-i\omega_kt}\tilde\rho(\omega_k)\int_0^t dt'\ e^{-\frac{\Gamma_{\text{a}}}{2}t'}e^{-i(\omega_{\text{a}}-\omega_k+\Delta_{\text{a}})t'}.
\end{equation}

\section{Energy density expectation value}\label{expec ener}

Following \cite{BialynickiBirula1998,Federico2023a}, we want to characterize the localization of the emitted single-photon state using the energy density operator
\begin{equation}
\hat{\mathcal{E}}_{\text{em}}(\vec x)=\frac{\varepsilon_0}{2}:\left(\vec{\hat E}^2(\vec x)+c^2\vec{\hat B}^2(\vec x)\right):,\label{local energy op}
\end{equation}
for which the expectation value for a single-photon state $\ket*{\vec\psi}$ in the Landau-Peierls (LP) representation (see Appendix \ref{LP and helicity}) reads \cite{BialynickiBirula1998,Federico2023a}
\begin{equation}\label{mean value}
\langle\hat{\mathcal{E}}_{\text{em}}(\vec x)\rangle=\hbar\abs{\Omega^{1/2}\vec \psi^{(h+)}(\vec x)}^2+\hbar\abs{\Omega^{1/2}\vec\psi^{(h-)}(\vec x)}^2.
\end{equation}
The exponents $^{(h\pm)}$ refer here to the positive and negative helicity parts of the LP representation (see Appendix \ref{LP and helicity}). This splitting into helicity components is an important feature of the result since it leads to the nonlocality of the energy density for all single-photon states \cite{Federico2023a} that we want to characterize for the spontaneous emission.

Using the results of the preceding section, we are going to express the state of the photon that is spontaneously emitted by the Lyman-$\alpha$ transition of the Hydrogen atom in the single-photon approximation, and then compute the associated expectation value of the electromagnetic energy density.
We consider here only the photon part of the state, i.e.,
\begin{equation}
\vec\psi_{\text{ph}}(r,\vartheta,\varphi)=\int_0^\infty dk\sum_{\lambda=\pm}c_{k,\lambda}(t)\vec\psi^{(\lambda)}_k(r,\vartheta,\varphi),
\end{equation}
where we use the abridged notation $\vec\psi^{(\lambda)}_k\equiv\vec\psi^{(\lambda)}_{k,J=1,M=m_e}$.
The positive and negative helicity parts of the state can be written as
\begin{equation}
\vec\psi_{\text{ph}}^{(h\pm)}(r,\vartheta,\varphi)=\int_0^\infty dk\frac{d_k(t)}{\sqrt{2}}\vec\psi^{(\pm)}_k(r,\vartheta,\varphi).
\end{equation}
Therefore, to compute the mean value, one needs to express
\begin{equation}
\abs{\Omega^{1/2}\vec\psi_{\text{ph}}^{(h\lambda)}(r,\vartheta,\varphi)}^2=\abs{\int_0^\infty dk\sqrt{\frac{\omega_k}{2}}d_k(t)\vec\psi^{(\lambda)}_k(r,\vartheta,\varphi)}^2\label{mean value term type}
\end{equation}
for both helicities. To do so, we use the expression of $\vec\psi_k^{(\lambda)}$ in terms of vector spherical harmonics and Bessel functions, as given in Appendix \ref{helicity functions}, to obtain
\begin{equation}
\abs{\Omega^{1/2}\vec\psi_{\text{ph}}^{(h\lambda)}}^2=\sum_{L=0}^2F_L(r,t)\vec Y_{1,m_e}^L(\vartheta,\varphi),
\end{equation}
where the $F_L$ functions are
\begin{subequations}
\begin{align}
F_0(r,t)&=i\sqrt{\frac{c}{3\pi}}\int_0^\infty dk\ k^{3/2}d_k(t)j_0(kr),\\
F_1(r,t)&=\lambda\sqrt{\frac{c}{2\pi}}\int_0^\infty dk\ k^{3/2}d_k(t)j_1(kr),\\
F_2(r,t)&=-i\sqrt{\frac{c}{6\pi}}\int_0^\infty dk\ k^{3/2}d_k(t)j_2(kr).
\end{align}
\end{subequations}
In order to simplify the expression of the mean value, we perform an asymptotic analysis of the $F_L$ functions for distances $r$ far from the atom. The details of this analysis are given in the Appendix \ref{asymptotic}. We obtain that the leading term is $F_1$ which behaves as
\begin{equation}
F_1(r,t)\underset{r\to\infty}{\sim}-i\lambda\left(\frac{2}{3}\right)^{11/2}\sqrt{\frac{\alpha^5}{c}}\frac{mc^2r_{\text{B}}^2}{\pi\hbar }\mathcal{T}(t)\frac{1}{r^3},
\end{equation}
where $\mathcal{T}(t)$ is a dimensionless function of time given in equation (\ref{time function}). The expectation value of the energy density becomes thus in the asymptotic limit of large $r$
\begin{equation}
\abs{\Omega^{1/2}\vec\psi_{\text{ph}}^{(h\lambda)}(\vec x)}^2\underset{r
\to\infty}{\sim}\abs*{F_1(r,t)}^2\vec Y_{1,m_e}^{1\star}\cdot\vec Y_{1,m_e}^1,
\end{equation}
which can be computed for different values of $m_e$. 
\begin{figure}
\includegraphics[height=5cm]{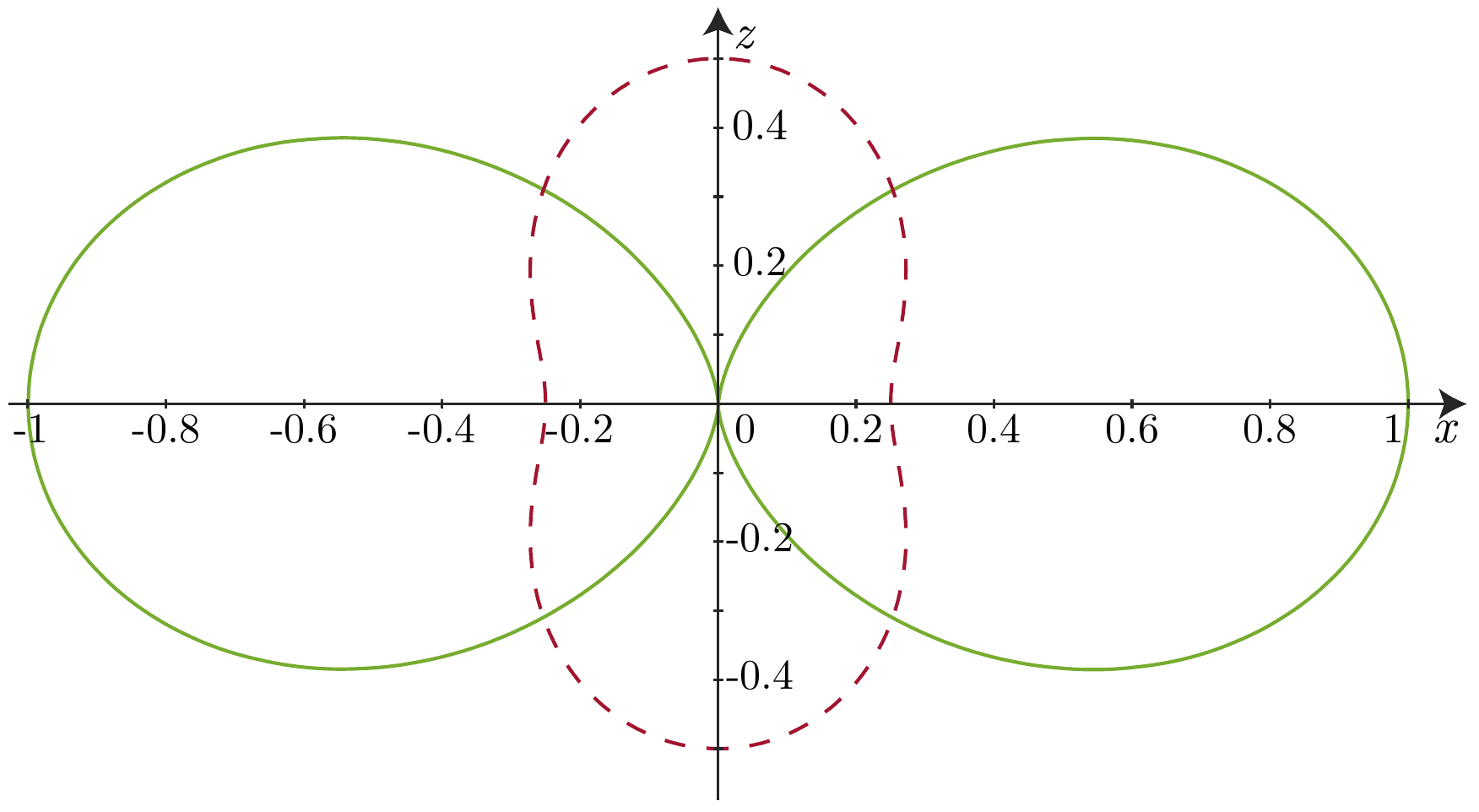}
\centering
\caption{Plot of the angular distribution $\gamma_{\abs*{m_e}}(\vartheta)$ of the photon state emitted by spontaneous emission of the Lyman-$\alpha$ transition of a Hydrogen atom, obtained in the asymptotic limit of large distances. The green solid line is the distribution for an excited state with $m_e=0$. The red dashed line is the distribution for an excited state with $\abs*{m_e}=1$.}
\label{angular dist}
\end{figure}
For $m_e=0$, the angular part reads
\begin{equation}
\vec Y_{1,0}^{1\star}\cdot\vec Y_{1,0}^1=\frac{3}{8\pi}\sin^2\vartheta,
\end{equation}
and for $\abs*{m_e}=1$
\begin{equation}
\vec Y_{1,\pm1}^{1\star}\cdot\vec Y_{1,\pm1}^1=\frac{3}{16\pi}\left(1+\cos^2\vartheta\right).
\end{equation}
In the end, the expectation value as a funtion of $m_e$ can be expressed as
\begin{equation}\label{final expect val}
\langle\hat{\mathcal{E}}_{\text{em}}\rangle=\left(\frac{2}{3}\right)^{10}\frac{m^2c^3r_{\text{B}}^4\alpha^5}{2\pi^3\hbar}\gamma_{\abs*{m_e}}(\vartheta)\abs{\mathcal{T}(t)}^2\frac{1}{r^6}+o\left(\frac{1}{r^6}\right),
\end{equation}
where the angular distribution is given by the functions $\gamma_{\abs*{m_e}}$ which read for the different values of $m_e$
\begin{equation}
\gamma_0(\vartheta)=\sin^2\vartheta,\quad\quad\quad
\gamma_1(\vartheta)=\frac{1}{4}(1+\cos^2\vartheta),
\end{equation}
and are represented in FIG. \ref{angular dist}. We point out that the angular distribution we have found matches that of a classical dipole that is either oscillating linearly for $\gamma_0$ or circularly for $\gamma_1$ \cite{Jackson1999,Steck2023}. We further remark that in the $m_e=0$ case, the leading term of the asymptotics is zero for $\vartheta=0$ due to the shape of the angular distribution. However, we emphasize that the spontaneous emission in this direction is not zero but features a faster falloff.

The result obtained in equation (\ref{final expect val}) exhibits an algebraic decrease for the energy density which is therefore far from the quasiexponential limit introduced in \cite{BialynickiBirula1998}, and cannot be attributed only to the blurring of the light cone as described in \cite{Shirokov1978}. This behavior is an illustration of the intrinsic nonlocality of the energy density of single-photon states, as shown for particular cases in \cite{BialynickiBirula2009} and for all single-photon states in \cite{Federico2023a}. It is also an other demonstration of the importance of multiphoton components that are predicted by the coupling to guarantee a causal emission \cite{Compagno1990}.

\section{Conclusion and outlook}

Under the assumption that the spontaneous emission of a Hydrogen atom produces a single-photon, we have determined the spatial nonlocality properties, and shown that at any given time the expectation value of the local energy density is not zero anywhere, and decreases asymptotically like a power $1/r^6$ as a function of the distance $r$ of the detector to the atom. We remark that this power law decrease is far from the maximal decrease, which is quasiexponential, as determined in \cite{BialynickiBirula1998}.
It was shown in \cite{Federico2023a} that all single-photon states have a nonzero expectation value of the energy in any finite volume. Our result provides a concrete illustration of the intrinsic nonlocal property of single photons through a calculation that does not only consider the field but its source as well, and it shows that the nonlaclity exists even before the photon state is completely emitted. From these results, one can conclude that the process of spontaneous emission cannot produce a strict single-photon state since it would contradict causality \cite{Berman2004,Dolce2006}. It is known that there are states for which the expectation values of all observables are localized in space \cite{Knight1961,Licht1963,Licht1966,DeBievre2006,DeBievre2007}. These states contain necessarily $n$-photon components for arbitrarily high $n$, with weights that can be small but nonzero. Coherent states are an example. An open question that should be addressed is the determination of the weights of the $n$-photon components in concrete emission processes, like spontaneous emission or in the production of photons at demand by the different techniques that are being developed experimentally.

\begin{acknowledgments}

We thank Jonas Lampart for fruitful discussions. This work was supported by the ``Investissements d'Avenir'' program, project ISITE-BFC/IQUINS (ANR-15-IDEX-03), QUACO-PRC (ANR-17-CE40-0007-01) and the EUR-EIPHI Graduate School (17-EURE-0002).
We also acknowledge support from the European Union's Horizon 2020 research and innovation program under the Marie Sklodowska-Curie Grant Agreement No. 765075 (LIMQUET) and from France 2030 (ANR-22-CMAS-0001) QuanTEdu-France. 
M. Federico also ackowledges funding from European Union’s Horizon Europe research and innovation programme under the project Quantum Secure Network Partnership (QSNP, grant agreement No 101114043).

\end{acknowledgments}

\appendix

\section{Position space representations and related operators}\label{LP and helicity}

To analyze the spatial properties of photons, one needs to use a position space representation of the states. This can be done using several equivalent choices \cite{Federico2023}. In this work, we have chosen to work with the Landau-Peierls (LP) representation which is constructed as follows
\begin{equation}
\vec\psi(\vec x)=\sqrt{\frac{\varepsilon_0}{2\hbar}}\left[\Omega^{1/2}\vec A(\vec x)-i\Omega^{-1/2}\vec E(\vec x)\right],\label{LP}
\end{equation}
where $\vec A$ is the classical potential vector in the Coulomb gauge and $\vec E$ is the classical electric field. The operators $\Omega^{\pm1/2}$ are self-adjoint operators constructed from the frequency operator $\Omega=c(-\Delta)^{1/2}$ for the Laplacian $\Delta$. These operators are well defined and Refs \cite{Federico2022,Federico2023,Federico2023a} give the more details of their construction. The LP representation is made in the Hilbert space
\begin{equation}
\mathcal{H}_{LP}=\left\{ \vec\psi\ \Big| \nabla\cdot\vec\psi=0, \bra*{\vec\psi}\ket*{\vec\psi'}_{LP}<\infty\right\},
\end{equation}
with the scalar product
\begin{equation}
\bra*{\vec\psi}\ket*{\vec\psi'}_{LP}=\int_{\mathbb{R}^3}d^3x \ \vec\psi^\star\cdot\vec\psi',
\end{equation}
and Maxwell's equations written in this representation read
\begin{equation}
i\frac{\partial\vec\psi}{\partial t}=\Omega\vec\psi,\quad\quad\quad \nabla\cdot\vec\psi=0.\label{Max eq}
\end{equation}
This Hilbert space can then be used to construct the bosonic Fock space of states
as
\begin{equation}
\mathbb{F}^{\mathfrak{B}}(\mathcal{H}_{LP})=\bigoplus_{n=0}^{\infty} \mathcal{H}_{LP}^{\otimes_{S} n},
\end{equation}
where $\mathcal{H}_{LP}^{\otimes_S n}$ is the symmetrized $n$-times tensor product of $\mathcal{H}_{LP}$ with itself.
Annihilation and creation operators directly constructed on arbitrary pulse-shaped functions can then be used to write states of the fields from the vacuum $\ket{\varnothing}$:
\begin{equation}\label{single photon state}
\hat B_{\vec\psi}\ket{\varnothing}=0,\quad\quad\quad
\hat B^\dag_{\vec\psi}\ket{\varnothing}=\ket*{\vec\psi}.
\end{equation}
They satisfy the general bosonic commutation relations \cite{Garrison2008,Fabre2020,Federico2023}
\begin{equation}
\left[ \hat B_{\vec\psi},\hat B^\dag_{\vec\psi'} \right]=\bra*{\vec\psi}\ket*{\vec\psi'}_{LP}.\label{comm rel LP}
\end{equation}
We emphasize that the definition is done for any function of $\mathcal{H}_{LP}$. As usual, one can also use any generalized basis like the circularly polarized plane waves or the helicity vector spherical harmonics introduced in Appendix \ref{helicity functions}, as we did in eq. (\ref{free elm}). Defining a basis allows also to express the electromagnetic field operators. Indeed, if we consider a basis $\{\vec\varphi_\kappa\}$, the classical LP function $\vec\psi(\vec x)$ can be expanded as
\begin{equation}
\vec\psi(\vec x)=\sum_\kappa \vec\varphi(\vec x)z_\kappa,
\end{equation}
with $z_\kappa=\bra*{\vec\varphi_\kappa}\ket*{\vec\psi}_{LP}$. Its quantum equivalent is thus 
\begin{equation}
\vec{\hat \Psi}(\vec x)=\sum_\kappa \vec\varphi(\vec x)\hat B_{\vec\varphi_\kappa},
\end{equation}
from which the real electromagnetic field observables are expressed as the following Hermitian operators
\begin{equation}
\vec{\hat A}(\vec x)=\sqrt{\frac{\hbar}{2\varepsilon_0}}\Omega^{-1/2}\left(\vec{\hat \Psi}(\vec x)+\vec{\hat \Psi}^\dag(\vec x)\right),\quad\quad\quad
\vec{\hat E}(\vec x)=i\sqrt{\frac{\hbar}{2\varepsilon_0}}\Omega^{1/2}\left(\vec{\hat \Psi}(\vec x)-\vec{\hat \Psi}^\dag(\vec x)\right).
\end{equation}
Another important tool that is used is the helicity operator, defined as
\begin{equation}
\Lambda=c\Omega^{-1}\nabla\times.
\end{equation}
For transverse fields, its spectrum is $\{\pm1\}$ and one can thus 
decompose any transverse field $\vec v$ into a sum of a positive- and
a negative-helicity part $\vec v^{(h\pm)}$,
\begin{equation}
\vec v=\vec v^{(h+)}+\vec v^{(h-)},
\end{equation}
where $\Lambda\vec v^{(h\pm)}=\pm\vec v^{(h\pm)}$. This splitting is an important property for the energy density mean value (\ref{mean value}) since it leads to the nonlocality of the energy density of all single-photon sates \cite{Federico2023a}.

\section{Helicity vector spherical harmonics}\label{helicity functions}

To compute the state of the emitted photon, we have used the basis of helicity vector spherical harmonics $\ket*{k,J,M,\lambda}\equiv\vec\psi_{k,J,M}^{(\lambda)}(\vec x)$ that are eigenfunctions of the following operators \cite{Moses1973,Varshalovich1988,Dai2012}
\begin{subequations}
\begin{align}
-\Delta\vec\psi_{k,J,M}^{(\lambda)}&=k^2\vec\psi_{k,J,M}^{(\lambda)}, \hspace{1.2cm} k\in(0,\infty),\\
\hat J^2\vec\psi_{k,J,M}^{(\lambda)}&=J(J+1)\vec\psi_{k,J,M}^{(\lambda)},\hspace{.2cm} J\in\{0,1,2,\dots\},\\
\hat J_3\vec\psi_{k,J,M}^{(\lambda)}&=M\vec\psi_{k,J,M}^{(\lambda)},\hspace{.99cm} M\in\{-J,\dots,J\},\\
\Lambda\vec\psi_{k,J,M}^{(\lambda)}&=\lambda\vec\psi_{k,J,M}^{(\lambda)}\hspace{1.55cm}\lambda=\pm1,
\end{align}
\end{subequations}
where 
\begin{subequations}
\begin{align}
\vec{\hat J}&=\vec{\hat L}+\vec{\hat S},\\
\vec{\hat L}&=\nabla\times\vec x = \vec x\times\nabla,\\
\vec{\hat S}&=\left( \hat S_1, \hat S_2, \hat S_3 \right)^T, \quad [\hat S_l]_{m,n}=-i\epsilon_{lmn},
\end{align}
\end{subequations}
where $l,m,n\in\{1,2,3\}$.
\noindent
The helicity vector spherical eigenfunctions can be written in spherical coordinates $(r,\vartheta,\varphi)$ as
\begin{align}\label{helicity vector sph harmonics}
\vec\psi_{k,J,M}^{(\lambda)}(r,\vartheta,\varphi)=\frac{i}{\sqrt{2}}&\Bigg[\sqrt{\frac{J+1}{2J+1}}\vec\psi_{k,J,M}^{J-1}(r,\vartheta,\varphi) - \sqrt{\frac{J}{2J+1}}\vec\psi_{k,J,M}^{J+1}(r,\vartheta,\varphi) - i \lambda \vec\psi_{k,J,M}^J(r,\vartheta,\varphi) \Bigg],
\end{align}
where the functions $\vec\psi_{k,J,M}^L$ can be expressed in terms of vector spherical harmonics $\vec Y_{J,M}^L$, i.e., eigenfunctions of $\hat J^2$, $\hat J_3$, $\hat L^2$ and $\hat S^2$ but not of $\Lambda$, and spherical Bessel functions $j_L(kr)$ as
\begin{equation}
\vec\psi_{k,J,M}^L(r,\vartheta,\varphi)=\sqrt{\frac{2}{\pi}}k\ j_L(kr)\vec Y_{J,M}^L(\vartheta,\varphi).
\end{equation}
We remark that by constructing the functions $\vec\psi_{k,J,M}^{(\lambda)}$ from $\vec\psi_{k,J,M}^L$ we obtain eigenfunctions of  $-\Delta$, $\hat J^2$, $\hat J_3$ and $\Lambda$ but not of $\hat L^2$ while $\vec\psi_{k,J,M}^L$ is an eigenfunction of  $-\Delta$, $\hat J^2$, $\hat J_3$ and $\hat L^2$ but not of $\Lambda$.

\noindent
The helicty spherical vector eigenfunctions $\{\vec\psi_{k,J,M}^{(\lambda)}\}$ are transverse, $\nabla\cdot\vec\psi_{k,J,M}^{(\lambda)}=0$, orthonormal 
\begin{align}
\int_0^\infty dr\ r^2 &\int_0^\pi d\vartheta \ \sin\vartheta \int_0^{2\pi} d\varphi \ \vec\psi_{k,J,M}^{\lambda\star}\cdot \vec\psi_{k',J',M'}^{\lambda'}=\delta(k-k')\delta_{J,J'}\delta_{M,M'}\delta_{\lambda,\lambda'},
\end{align}
and form a complete set of the subspace of transverse field since the set of function $\{\vec\psi_{k,J,M}^{J-1},\vec\psi_{k,J,M}^J,\vec\psi_{k,J,M}^{J+1}\}$ from which they are defined is a complete set with
\begin{align}
\int_0^\infty d^3k&\sum_{J,M,L}\left[\vec\psi_{k,J,M}^L(r',\vartheta',\varphi')\right]_l\left[\vec\psi_{k,J,M}^{L\star}(r,\vartheta,\varphi)\right]_n=\frac{1}{r^2\sin\vartheta}\delta_{l,n}\delta(r-r')\delta(\vartheta-\vartheta')\delta(\varphi-\varphi').
\end{align}
The last equation is a consequence of the completeness of the Bessel functions $j_L(kr)$ and vector spherical harmonics $\vec Y_{J,M}^L$ \cite{Varshalovich1988}.

\section{Asymptotic analysis of radial integrals}\label{asymptotic}

In this appendix, we compute the asymptotic behavior of the radial integrals that appear in the calculation of the energy density of a spontaneously emitted photon from the Lyman-$\alpha$ transition of a Hydrogen atom. The integrals have the following forms
\begin{widetext}
\begin{subequations}
\begin{align}
F_0(r,t)&=\left(\frac{2}{3}\right)^3\sqrt{\frac{\alpha^5}{c}}\frac{mc^2}{\pi\hbar r_\text{B}}\int_0^\infty dq\frac{e^{-iqp}-e^{-i(\Omega_{\text{a}}-i\tilde\Gamma_{\text{a}})t}}{A+i(B-q)}\frac{q^2}{\left(1+q^2\right)^2}j_0(qr'),\\
F_1(r,t)&=-i\lambda\left(\frac{2}{3}\right)^{5/2}\sqrt{\frac{\alpha^5}{c}}\frac{mc^2}{\pi\hbar r_\text{B}}\int_0^\infty dq\frac{e^{-iqp}-e^{-i(\Omega_{\text{a}}-i\tilde\Gamma_{\text{a}})t}}{A+i(B-q)}\frac{q^2}{\left(1+q^2\right)^2}j_1(qr'),\\
F_2(r,t)&=-\left(\frac{2}{3}\right)^3\sqrt{\frac{\alpha^5}{2c}}\frac{mc^2}{\pi\hbar r_\text{B}}\int_0^\infty dq\frac{e^{-iqp}-e^{-i(\Omega_{\text{a}}-i\tilde\Gamma_{\text{a}})t}}{A+i(B-q)}\frac{q^2}{\left(1+q^2\right)^2}\left( \frac{3}{qr'}j_1(qr')-j_0(qr') \right),
\end{align}
\end{subequations}
where we have introduced the dimensionless parameters
\begin{subequations}
\label{unitless par}
\begin{align}
A=\frac{\tilde\Gamma_{\text{a}}}{cK}=\frac{\Gamma_{\text{a}}}{2cK},\quad\quad\quad B=\frac{\Omega_{\text{a}}}{cK}=\frac{\omega_{\text{a}}+\Delta_\text{a}}{cK},
\end{align}
\begin{align}
q=\frac{k}{K},\quad\quad\quad p=cKt, \quad\quad\quad r'=Kr.
\end{align}
\end{subequations} 
To obtain the asymptotic behavior of the $F_L$ integrals, and due to the expression of the spherical Bessel functions 
\begin{equation}
j_0(qr')=\frac{\sin(qr')}{qr'},\quad\quad\quad j_1(qr')=\frac{\sin(qr')}{q^2r'^2}-\frac{\cos(qr')}{qr'},
\end{equation}
one needs to analyze integrals of the form
\begin{equation}\label{cos sin int}
\int_0^\infty dqR(q)\sin(qr'),\quad\quad\quad
\int_0^\infty dqR(q)\cos(qr').
\end{equation}
Here, $R(q)$ is a function with the properties that for some $N_s\geq0$ even or $N_c\geq0$ odd,
\begin{subequations}
\begin{align}
\left.\frac{d^nR(q)}{dq^n}\right|_{q=0}&=\left\{
\begin{array}{ll}
0\quad& \text{for $n<N_{c,s}$, $n$ same parity as $N_{c,s}$},\\
\text{cst}\neq0\quad& \text{for $n= N_{c,s}$, $n$ same parity as $N_{c,s}$},
\end{array}
\right.\\
\lim_{q\to\infty}\frac{d^nR(q)}{dq^n}&=0,
\end{align}
\end{subequations}
and
\begin{equation}\label{condition for RL lemma}
\int_0^\infty dq \abs{\frac{d^nR(q)}{dq^n}}<\infty, \quad \text{for all $n<N_{c,s}$}.
\end{equation}
The indices $c,s$, are needed to treat the cosine and sine integrals in (\ref{cos sin int}), respectively.
Using these properties and some integrations by parts the sine intergral becomes
\begin{subequations}
\begin{align}
\int_0^\infty dqR(q)\sin(qr')&=-\frac{1}{r'}\Big[R(q)\cos(qr')\Big]_0^\infty+\frac{1}{r'}\int_0^\infty dq\frac{dR(q)}{dq}\cos(qr')\\
&=\frac{1}{r'}R(0)-\frac{1}{r'}\int_0^\infty dq\frac{dR(q)}{dq}\cos(qr')\\
&=\frac{1}{r'}R(0)+o\left(\frac{1}{r'}\right).\label{sine asympt gen}
\end{align}
\end{subequations}
The last equality is obtained by the Riemann-Lebesgue lemma \cite{Appel2007} which states that 
\begin{equation}
\abs{\int_0^\infty dq\frac{dR(q)}{dq}\cos(qr')}\underset{r'\rightarrow+\infty}{\longrightarrow}0,
\end{equation} 
if $\int_0^\infty dq \abs{\frac{dR(q)}{dq}}<\infty$, which is true according to (\ref{condition for RL lemma}). Therefore, the first term of (\ref{sine asympt gen}) gives the asymptotics for large $r'$ if $N_s=0$, otherwise, one has to iterate the integration by parts as
\begin{subequations}
\begin{align}
\int_0^\infty dqR(q)\sin(qr')&=\frac{1}{r'^2}\left[\frac{dR}{dq}\sin(qr')\right]_0^\infty-\frac{1}{r'^2}\int_0^\infty dq\frac{d^2R(q)}{dq^2}\sin(qr')\\
&=-\frac{1}{r'^2}\int_0^\infty dq\frac{d^2R(q)}{dq^2}\sin(qr')\\
&=\frac{1}{r'^3}\left[\frac{d^2R}{dq^2}\cos(qr')\right]_0^\infty+\frac{1}{r'^3}\int_0^\infty dq\frac{d^3R(q)}{dq^3}\cos(qr')\\
&=\frac{1}{r'^3}\left.\frac{d^2R}{dq^2}\right|_{q=0}+\frac{1}{r'^3}\int_0^\infty dq\frac{d^3R(q)}{dq^3}\cos(qr')\\
&=\frac{1}{r'^3}\left.\frac{d^2R}{dq^2}\right|_{q=0}+o\left(\frac{1}{r'^3}\right),
\end{align}
\end{subequations}
and again, the first term gives the asymptotics by the Riemann-Lebesgue lemma if $N_s=2$. Otherwise, the same process can be continued until one reaches $N_s$.
A similar process can be done for the cosine integral and gives after one step
\begin{subequations}
\label{cos step}
\begin{align}
\int_0^\infty dqR(q)\cos(qr')&=-\frac{1}{r'^2}\left.\frac{dR}{dq}\right|_{q=0}+\frac{1}{r'^2}\int_0^\infty dq\frac{d^2R(q)}{dq^2}\cos(qr')\\
&=-\frac{1}{r'^2}\left.\frac{dR}{dq}\right|_{q=0}+o\left(\frac{1}{r'^2}\right),
\end{align}
\end{subequations}
by the Riemann-Lebesgue lemma. This step is enough if $N_c=1$ but others can be performed if not.

\noindent
To apply this technique, we now need the concrete form of the functions $R(q)$ and thus we rewrite the $F_L$ integrals as
\begin{subequations}
\begin{align}
F_0&=\left(\frac{2}{3}\right)^3\sqrt{\frac{\alpha^5}{c}}\frac{mc^2}{\pi\hbar r_\text{B}}\frac{1}{r'}\int_0^\infty dq\left(R^{(1)}_+(q)+R^{(1)}_-(q)\right)\sin(qr'),\\
F_1&=-i\lambda\left(\frac{2}{3}\right)^{5/2}\sqrt{\frac{\alpha^5}{c}}\frac{mc^2}{\pi\hbar r_\text{B}}\Bigg\{\frac{1}{r'^2}\int_0^\infty dq\left(R^{(0)}_+(q)+R^{(0)}_-(q)\right)\sin(qr')-\frac{1}{r'}\int_0^\infty dq\left(R^{(1)}_+(q)+R^{(1)}_-(q)\right)\cos(qr')\Bigg\},
\end{align}
\end{subequations}
where the $R$ functions are
\begin{subequations}
\begin{align}
R_+^{(1)}(q)=\frac{e^{-iqp}}{A+i(B-q)}\frac{q}{\left(1+q^2\right)^2}, &\quad\quad\quad R_-^{(1)}(q)=-\frac{e^{-i(\Omega_{\text{a}}-i\tilde\Gamma_{\text{a}})t}}{A+i(B-q)}\frac{q}{\left(1+q^2\right)^2},\\
R_+^{(0)}(q)=\frac{e^{-iqp}}{A+i(B-q)}\frac{1}{\left(1+q^2\right)^2}, &\quad\quad\quad R_-^{(0)}(q)=-\frac{e^{-i(\Omega_{\text{a}}-i\tilde\Gamma_{\text{a}})t}}{A+i(B-q)}\frac{1}{\left(1+q^2\right)^2}.
\end{align}
\end{subequations}
The integral $F_2$ can be deduced rapidly from the other two as we will discuss later. The asymptotics for $F_0$ and $F_1$ can be computed using the iterative process and we obtain
\begin{subequations}
\label{asymptotics with sum}
\begin{align}
F_0(r,t)\underset{r'\to\infty}{\sim}&\left(\frac{2}{3}\right)^3\sqrt{\frac{\alpha^5}{c}}\frac{mc^2}{\pi\hbar r_\text{B}}\frac{1}{r'^4}\left(\left.\frac{d^2R_+^{(1)}}{dq^2}\right|_{q=0}+\left.\frac{d^2R_-^{(1)}}{dq^2}\right|_{q=0}\right),\\
F_1(r,t)\underset{r'\to\infty}{\sim}&-i\lambda\left(\frac{2}{3}\right)^{5/2}\sqrt{\frac{\alpha^5}{c}}\frac{mc^2}{\pi\hbar r_\text{B}}\frac{1}{r'^3}\left(R_+^{(0)}(0)+R_-^{(0)}(0)+\left.\frac{d^2R_+^{(1)}}{dq^2}\right|_{q=0}+\left.\frac{d^2R_-^{(1)}}{dq^2}\right|_{q=0}\right),\label{asymptotics with sum1}
\end{align}
\end{subequations}
which can be further simplified using
\begin{subequations}
\label{R in zero}
\begin{align}
R^{(0)}_+(q=0)&=\frac{1}{A+iB},\hspace{4cm}
R^{(0)}_-(q=0)=-\frac{e^{-i(\Omega_{\text{a}}-i\tilde\Gamma_{\text{a}})t}}{A+iB},\\
\left.\frac{d^2R_+^{(1)}}{dq^2}\right|_{q=0}&=\frac{i}{(A+iB)^2}-\frac{ip}{A+iB},\hspace{1.95cm}
\left.\frac{d^2R^{(1)}_-}{dq^2}\right|_{q=0}=-i\frac{e^{-i(\Omega_{\text{a}}-i\tilde\Gamma_{\text{a}})t}}{(A+iB)^2}.
\end{align}
\end{subequations}
Since $F_0\sim F_1/r'$, we compute only $F_1$ which will dominate for large $r'$. We obtain then the result used in Section \ref{expec ener} by inserting (\ref{R in zero}) in (\ref{asymptotics with sum1})
\begin{equation}
F_1(r',t)\underset{r'\to\infty}{\sim}-i\lambda\left(\frac{2}{3}\right)^{5/2}\sqrt{\frac{\alpha^5}{c}}\frac{mc^2}{\pi\hbar r_\text{B}}\mathcal{T}(t)\frac{1}{r'^3},
\end{equation}
where the dimensionless function of time reads
\begin{equation}\label{time function}
\mathcal{T}(t)=\frac{1}{A+iB}\left(1-e^{-i(\Omega_{\text{a}}-i\tilde\Gamma_{\text{a}})t}-2icKt+i\frac{1-e^{-i(\Omega_{\text{a}}-i\tilde\Gamma_{\text{a}})t}}{A+iB}\right).
\end{equation}
Regarding $F_2$ we recall its expression 
\begin{equation}
F_2(r,t)=-\left(\frac{2}{3}\right)^3\sqrt{\frac{\alpha^5}{2c}}\frac{mc^2}{\pi\hbar r_\text{B}}\int_0^\infty dq\frac{e^{-iqp}-e^{-i(\Omega_{\text{a}}-i\tilde\Gamma_{\text{a}})t}}{A+i(B-q)}\frac{q^2}{\left(1+q^2\right)^2}\left( \frac{3}{qr'}j_1(qr')-j_0(qr') \right),
\end{equation}
from which we see that the term involving $j_0$ will behave as $F_0$ and can thus be neglected. For the term involving $j_1$, one has to be more cautious since it does not have the same form as what we had for $F_1$. It yields
\begin{subequations}
\begin{align}
F_2(r',t)\underset{r'\to\infty}{\sim}&-\left(\frac{2}{3}\right)^2\sqrt{\frac{2\alpha^5}{c}}\frac{mc^2}{\pi\hbar r_\text{B}}\frac{1}{r'}\int_0^\infty dq\frac{e^{-iqp}-e^{-i(\Omega_{\text{a}}-i\tilde\Gamma_{\text{a}})t}}{A+i(B-q)}\frac{q}{\left(1+q^2\right)^2}j_1(qr')\\
&=-\left(\frac{2}{3}\right)^2\sqrt{\frac{2\alpha^5}{c}}\frac{mc^2}{\pi\hbar r_\text{B}}\left\{\frac{1}{r'^3}\int_0^\infty dq\frac{e^{-iqp}-e^{-i(\Omega_{\text{a}}-i\tilde\Gamma_{\text{a}})t}}{A+i(B-q)}\frac{1}{q\left(1+q^2\right)^2}\sin(qr')\right.\nonumber\\&\:\:\:\:\:\:\left.-\frac{1}{r'^2}\int_0^\infty dq\frac{e^{-iqp}-e^{-i(\Omega_{\text{a}}-i\tilde\Gamma_{\text{a}})t}}{A+i(B-q)}\frac{1}{\left(1+q^2\right)^2}\cos(qr')\right\},
\end{align}
\end{subequations}
\end{widetext}
from which one can already see that $F_2$ will decrease faster than $F_1$. Indeed, the first term scales like $1/r'^3$ before making the iterative process to analyze the integral. Moreover, the function of $q$ to integrate does not yield a constant when $q\rightarrow0$ meaning that the iterative process must be done more than one time. Thus, this term will decrease faster than $1/r'^3$ and can be neglected. The second term involves a cosine for which the first iterative step (\ref{cos step}) already scales as $1/r'^2$, so that the full term would scale like $1/r'^4$, decreasing faster than $F_1$. The leading term that is needed to compute the mean value is thus given by $F_1$ and if one reinstates the units using (\ref{unitless par}), it yields
\begin{equation}
F_1(r,t)\underset{r\to\infty}{\sim}-i\lambda\left(\frac{2}{3}\right)^{11/2}\sqrt{\frac{\alpha^5}{c}}\frac{mc^2r_{\text{B}}^2}{\pi\hbar }\mathcal{T}(t)\frac{1}{r^3}.
\end{equation}

\bibliography{sp-emission}

\begin{thebibliography}{53}%
\makeatletter
\providecommand \@ifxundefined [1]{%
 \@ifx{#1\undefined}
}%
\providecommand \@ifnum [1]{%
 \ifnum #1\expandafter \@firstoftwo
 \else \expandafter \@secondoftwo
 \fi
}%
\providecommand \@ifx [1]{%
 \ifx #1\expandafter \@firstoftwo
 \else \expandafter \@secondoftwo
 \fi
}%
\providecommand \natexlab [1]{#1}%
\providecommand \enquote  [1]{``#1''}%
\providecommand \bibnamefont  [1]{#1}%
\providecommand \bibfnamefont [1]{#1}%
\providecommand \citenamefont [1]{#1}%
\providecommand \href@noop [0]{\@secondoftwo}%
\providecommand \href [0]{\begingroup \@sanitize@url \@href}%
\providecommand \@href[1]{\@@startlink{#1}\@@href}%
\providecommand \@@href[1]{\endgroup#1\@@endlink}%
\providecommand \@sanitize@url [0]{\catcode `\\12\catcode `\$12\catcode
  `\&12\catcode `\#12\catcode `\^12\catcode `\_12\catcode `\%12\relax}%
\providecommand \@@startlink[1]{}%
\providecommand \@@endlink[0]{}%
\providecommand \url  [0]{\begingroup\@sanitize@url \@url }%
\providecommand \@url [1]{\endgroup\@href {#1}{\urlprefix }}%
\providecommand \urlprefix  [0]{URL }%
\providecommand \Eprint [0]{\href }%
\providecommand \doibase [0]{https://doi.org/}%
\providecommand \selectlanguage [0]{\@gobble}%
\providecommand \bibinfo  [0]{\@secondoftwo}%
\providecommand \bibfield  [0]{\@secondoftwo}%
\providecommand \translation [1]{[#1]}%
\providecommand \BibitemOpen [0]{}%
\providecommand \bibitemStop [0]{}%
\providecommand \bibitemNoStop [0]{.\EOS\space}%
\providecommand \EOS [0]{\spacefactor3000\relax}%
\providecommand \BibitemShut  [1]{\csname bibitem#1\endcsname}%
\let\auto@bib@innerbib\@empty
\bibitem [{\citenamefont {Weisskopf}\ and\ \citenamefont
  {Wigner}(1930)}]{Weisskopf1930}%
  \BibitemOpen
  \bibfield  {author} {\bibinfo {author} {\bibfnamefont {V.}~\bibnamefont
  {Weisskopf}}\ and\ \bibinfo {author} {\bibfnamefont {E.}~\bibnamefont
  {Wigner}},\ }\href {https://doi.org/10.1007/bf01336768} {\bibfield  {journal}
  {\bibinfo  {journal} {Zeitschrift f{\"u}r Physik}\ }\textbf {\bibinfo
  {volume} {63}},\ \bibinfo {pages} {54} (\bibinfo {year} {1930})}\BibitemShut
  {NoStop}%
\bibitem [{\citenamefont {Spohn}(2009)}]{Spohn2009}%
  \BibitemOpen
  \bibfield  {author} {\bibinfo {author} {\bibfnamefont {H.}~\bibnamefont
  {Spohn}},\ }\href@noop {} {\emph {\bibinfo {title} {Dynamics of Charged
  Particles and Their Radiation Field}}}\ (\bibinfo  {publisher} {Cambridge
  University Press},\ \bibinfo {year} {2009})\BibitemShut {NoStop}%
\bibitem [{\citenamefont {Novotny}\ and\ \citenamefont
  {Hecht}(2012)}]{Novotny2012}%
  \BibitemOpen
  \bibfield  {author} {\bibinfo {author} {\bibfnamefont {L.}~\bibnamefont
  {Novotny}}\ and\ \bibinfo {author} {\bibfnamefont {B.}~\bibnamefont
  {Hecht}},\ }\href {https://doi.org/10.1017/cbo9780511794193} {\emph {\bibinfo
  {title} {Principles of Nano-Optics}}}\ (\bibinfo  {publisher} {Cambridge
  University Press},\ \bibinfo {year} {2012})\BibitemShut {NoStop}%
\bibitem [{\citenamefont {Dereziński}\ and\ \citenamefont
  {Früboes}(2002)}]{Derezinski2002}%
  \BibitemOpen
  \bibfield  {author} {\bibinfo {author} {\bibfnamefont {J.}~\bibnamefont
  {Dereziński}}\ and\ \bibinfo {author} {\bibfnamefont {R.}~\bibnamefont
  {Früboes}},\ }\href {https://doi.org/10.1016/s0034-4877(02)80070-2}
  {\bibfield  {journal} {\bibinfo  {journal} {Reports on Mathematical Physics}\
  }\textbf {\bibinfo {volume} {50}},\ \bibinfo {pages} {433} (\bibinfo {year}
  {2002})}\BibitemShut {NoStop}%
\bibitem [{\citenamefont {Moses}(1973)}]{Moses1973}%
  \BibitemOpen
  \bibfield  {author} {\bibinfo {author} {\bibfnamefont {H.~E.}\ \bibnamefont
  {Moses}},\ }\href {https://doi.org/10.1103/physreva.8.1710} {\bibfield
  {journal} {\bibinfo  {journal} {Physical Review A}\ }\textbf {\bibinfo
  {volume} {8}},\ \bibinfo {pages} {1710} (\bibinfo {year} {1973})}\BibitemShut
  {NoStop}%
\bibitem [{\citenamefont {Moses}(1975{\natexlab{a}})}]{Moses1975}%
  \BibitemOpen
  \bibfield  {author} {\bibinfo {author} {\bibfnamefont {H.~E.}\ \bibnamefont
  {Moses}},\ }\href {https://doi.org/10.1103/physreva.11.2196} {\bibfield
  {journal} {\bibinfo  {journal} {Physical Review A}\ }\textbf {\bibinfo
  {volume} {11}},\ \bibinfo {pages} {2196} (\bibinfo {year}
  {1975}{\natexlab{a}})}\BibitemShut {NoStop}%
\bibitem [{\citenamefont {Moses}(1975{\natexlab{b}})}]{Moses1975a}%
  \BibitemOpen
  \bibfield  {author} {\bibinfo {author} {\bibfnamefont {H.~E.}\ \bibnamefont
  {Moses}},\ }\href {https://doi.org/10.1103/physreva.11.387} {\bibfield
  {journal} {\bibinfo  {journal} {Physical Review A}\ }\textbf {\bibinfo
  {volume} {11}},\ \bibinfo {pages} {387} (\bibinfo {year}
  {1975}{\natexlab{b}})}\BibitemShut {NoStop}%
\bibitem [{\citenamefont {Grimm}\ and\ \citenamefont
  {Ernst}(1974)}]{Grimm1974}%
  \BibitemOpen
  \bibfield  {author} {\bibinfo {author} {\bibfnamefont {E.}~\bibnamefont
  {Grimm}}\ and\ \bibinfo {author} {\bibfnamefont {V.}~\bibnamefont {Ernst}},\
  }\href {https://doi.org/10.1088/0305-4470/7/13/020} {\bibfield  {journal}
  {\bibinfo  {journal} {Journal of Physics A: Mathematical, Nuclear and
  General}\ }\textbf {\bibinfo {volume} {7}},\ \bibinfo {pages} {1664}
  (\bibinfo {year} {1974})}\BibitemShut {NoStop}%
\bibitem [{\citenamefont {Grimm}\ and\ \citenamefont
  {Ernst}(1975)}]{Grimm1975}%
  \BibitemOpen
  \bibfield  {author} {\bibinfo {author} {\bibfnamefont {E.}~\bibnamefont
  {Grimm}}\ and\ \bibinfo {author} {\bibfnamefont {V.}~\bibnamefont {Ernst}},\
  }\href {https://doi.org/10.1007/bf01434039} {\bibfield  {journal} {\bibinfo
  {journal} {Zeitschrift f{\"u}r Physik A Atoms and Nuclei}\ }\textbf {\bibinfo
  {volume} {274}},\ \bibinfo {pages} {293} (\bibinfo {year}
  {1975})}\BibitemShut {NoStop}%
\bibitem [{\citenamefont {Seke}\ and\ \citenamefont
  {Herfort}(1988)}]{Seke1988}%
  \BibitemOpen
  \bibfield  {author} {\bibinfo {author} {\bibfnamefont {J.}~\bibnamefont
  {Seke}}\ and\ \bibinfo {author} {\bibfnamefont {W.~N.}\ \bibnamefont
  {Herfort}},\ }\href {https://doi.org/10.1103/physreva.38.833} {\bibfield
  {journal} {\bibinfo  {journal} {Physical Review A}\ }\textbf {\bibinfo
  {volume} {38}},\ \bibinfo {pages} {833} (\bibinfo {year} {1988})}\BibitemShut
  {NoStop}%
\bibitem [{\citenamefont {Knight}(1961)}]{Knight1961}%
  \BibitemOpen
  \bibfield  {author} {\bibinfo {author} {\bibfnamefont {J.~M.}\ \bibnamefont
  {Knight}},\ }\href {https://doi.org/10.1063/1.1703731} {\bibfield  {journal}
  {\bibinfo  {journal} {Journal of Mathematical Physics}\ }\textbf {\bibinfo
  {volume} {2}},\ \bibinfo {pages} {459} (\bibinfo {year} {1961})}\BibitemShut
  {NoStop}%
\bibitem [{\citenamefont {Hegerfeldt}(1974)}]{Hegerfeldt1974}%
  \BibitemOpen
  \bibfield  {author} {\bibinfo {author} {\bibfnamefont {G.~C.}\ \bibnamefont
  {Hegerfeldt}},\ }\href {https://doi.org/10.1103/physrevd.10.3320} {\bibfield
  {journal} {\bibinfo  {journal} {Physical Review D}\ }\textbf {\bibinfo
  {volume} {10}},\ \bibinfo {pages} {3320} (\bibinfo {year}
  {1974})}\BibitemShut {NoStop}%
\bibitem [{\citenamefont {Hegerfeldt}(1994)}]{Hegerfeldt1994}%
  \BibitemOpen
  \bibfield  {author} {\bibinfo {author} {\bibfnamefont {G.~C.}\ \bibnamefont
  {Hegerfeldt}},\ }\href {https://doi.org/10.1103/physrevlett.72.596}
  {\bibfield  {journal} {\bibinfo  {journal} {Physical Review Letters}\
  }\textbf {\bibinfo {volume} {72}},\ \bibinfo {pages} {596} (\bibinfo {year}
  {1994})}\BibitemShut {NoStop}%
\bibitem [{\citenamefont {Hegerfeldt}(1997)}]{Hegerfeldt1997}%
  \BibitemOpen
  \bibfield  {author} {\bibinfo {author} {\bibfnamefont {G.~C.}\ \bibnamefont
  {Hegerfeldt}},\ }\href@noop {} {\bibfield  {journal} {\bibinfo  {journal}
  {arXiv:quant-ph/9707016}\ } (\bibinfo {year} {1997})}\BibitemShut {NoStop}%
\bibitem [{\citenamefont {Hegerfeldt}(1998)}]{Hegerfeldt1998}%
  \BibitemOpen
  \bibfield  {author} {\bibinfo {author} {\bibfnamefont {G.~C.}\ \bibnamefont
  {Hegerfeldt}},\ }in\ \href {https://doi.org/10.1007/bfb0106784} {\emph
  {\bibinfo {booktitle} {Irreversibility and Causality Semigroups and Rigged
  Hilbert Spaces}}}\ (\bibinfo  {publisher} {Springer Berlin Heidelberg},\
  \bibinfo {year} {1998})\ pp.\ \bibinfo {pages} {238--245}\BibitemShut
  {NoStop}%
\bibitem [{\citenamefont {Bia{\l}ynicki-Birula}(1998)}]{BialynickiBirula1998}%
  \BibitemOpen
  \bibfield  {author} {\bibinfo {author} {\bibfnamefont {I.}~\bibnamefont
  {Bia{\l}ynicki-Birula}},\ }\href
  {https://doi.org/10.1103/physrevlett.80.5247} {\bibfield  {journal} {\bibinfo
   {journal} {Physical Review Letters}\ }\textbf {\bibinfo {volume} {80}},\
  \bibinfo {pages} {5247} (\bibinfo {year} {1998})}\BibitemShut {NoStop}%
\bibitem [{\citenamefont {Bia{\l}ynicki-Birula}\ and\ \citenamefont
  {Bia{\l}ynicka-Birula}(2009)}]{BialynickiBirula2009}%
  \BibitemOpen
  \bibfield  {author} {\bibinfo {author} {\bibfnamefont {I.}~\bibnamefont
  {Bia{\l}ynicki-Birula}}\ and\ \bibinfo {author} {\bibfnamefont
  {Z.}~\bibnamefont {Bia{\l}ynicka-Birula}},\ }\href
  {https://doi.org/10.1103/physreva.79.032112} {\bibfield  {journal} {\bibinfo
  {journal} {Physical Review A}\ }\textbf {\bibinfo {volume} {79}},\ \bibinfo
  {pages} {032112} (\bibinfo {year} {2009})}\BibitemShut {NoStop}%
\bibitem [{\citenamefont {Federico}\ and\ \citenamefont
  {Jauslin}(2023{\natexlab{a}})}]{Federico2023a}%
  \BibitemOpen
  \bibfield  {author} {\bibinfo {author} {\bibfnamefont {M.}~\bibnamefont
  {Federico}}\ and\ \bibinfo {author} {\bibfnamefont {H.~R.}\ \bibnamefont
  {Jauslin}},\ }\href {https://doi.org/10.1103/physreva.108.043720} {\bibfield
  {journal} {\bibinfo  {journal} {Physical Review A}\ }\textbf {\bibinfo
  {volume} {108}},\ \bibinfo {pages} {043720} (\bibinfo {year}
  {2023}{\natexlab{a}})}\BibitemShut {NoStop}%
\bibitem [{\citenamefont {Berman}(2004)}]{Berman2004}%
  \BibitemOpen
  \bibfield  {author} {\bibinfo {author} {\bibfnamefont {P.~R.}\ \bibnamefont
  {Berman}},\ }\href {https://doi.org/10.1103/physreva.69.022101} {\bibfield
  {journal} {\bibinfo  {journal} {Physical Review A}\ }\textbf {\bibinfo
  {volume} {69}},\ \bibinfo {pages} {022101} (\bibinfo {year}
  {2004})}\BibitemShut {NoStop}%
\bibitem [{\citenamefont {Dolce}\ \emph {et~al.}(2006)\citenamefont {Dolce},
  \citenamefont {Passante},\ and\ \citenamefont {Persico}}]{Dolce2006}%
  \BibitemOpen
  \bibfield  {author} {\bibinfo {author} {\bibfnamefont {I.}~\bibnamefont
  {Dolce}}, \bibinfo {author} {\bibfnamefont {R.}~\bibnamefont {Passante}},\
  and\ \bibinfo {author} {\bibfnamefont {F.}~\bibnamefont {Persico}},\ }\href
  {https://doi.org/10.1016/j.physleta.2006.02.018} {\bibfield  {journal}
  {\bibinfo  {journal} {Physics Letters A}\ }\textbf {\bibinfo {volume}
  {355}},\ \bibinfo {pages} {152} (\bibinfo {year} {2006})}\BibitemShut
  {NoStop}%
\bibitem [{\citenamefont {Gulla}\ and\ \citenamefont
  {Skaar}(2021)}]{Gulla2021}%
  \BibitemOpen
  \bibfield  {author} {\bibinfo {author} {\bibfnamefont {J.}~\bibnamefont
  {Gulla}}\ and\ \bibinfo {author} {\bibfnamefont {J.}~\bibnamefont {Skaar}},\
  }\href {https://doi.org/10.1103/physrevlett.126.073601} {\bibfield  {journal}
  {\bibinfo  {journal} {Physical Review Letters}\ }\textbf {\bibinfo {volume}
  {126}},\ \bibinfo {pages} {073601} (\bibinfo {year} {2021})}\BibitemShut
  {NoStop}%
\bibitem [{\citenamefont {Ryen}\ \emph {et~al.}(2022)\citenamefont {Ryen},
  \citenamefont {Gulla},\ and\ \citenamefont {Skaar}}]{Ryen2022}%
  \BibitemOpen
  \bibfield  {author} {\bibinfo {author} {\bibfnamefont {K.}~\bibnamefont
  {Ryen}}, \bibinfo {author} {\bibfnamefont {J.}~\bibnamefont {Gulla}},\ and\
  \bibinfo {author} {\bibfnamefont {J.}~\bibnamefont {Skaar}},\ }\bibfield
  {journal} {\bibinfo  {journal} {International Journal of Theoretical
  Physics}\ }\textbf {\bibinfo {volume} {61}},\ \href
  {https://doi.org/10.1007/s10773-022-05133-7} {10.1007/s10773-022-05133-7}
  (\bibinfo {year} {2022})\BibitemShut {NoStop}%
\bibitem [{\citenamefont {Gulla}\ \emph {et~al.}(2023)\citenamefont {Gulla},
  \citenamefont {Ryen},\ and\ \citenamefont {Skaar}}]{Gulla2023}%
  \BibitemOpen
  \bibfield  {author} {\bibinfo {author} {\bibfnamefont {J.}~\bibnamefont
  {Gulla}}, \bibinfo {author} {\bibfnamefont {K.}~\bibnamefont {Ryen}},\ and\
  \bibinfo {author} {\bibfnamefont {J.}~\bibnamefont {Skaar}},\ }\href
  {https://doi.org/10.1103/physreva.108.063708} {\bibfield  {journal} {\bibinfo
   {journal} {Physical Review A}\ }\textbf {\bibinfo {volume} {108}},\ \bibinfo
  {pages} {063708} (\bibinfo {year} {2023})}\BibitemShut {NoStop}%
\bibitem [{\citenamefont {Biswas}\ \emph {et~al.}(1990)\citenamefont {Biswas},
  \citenamefont {Compagno}, \citenamefont {Palma}, \citenamefont {Passante},\
  and\ \citenamefont {Persico}}]{Biswas1990}%
  \BibitemOpen
  \bibfield  {author} {\bibinfo {author} {\bibfnamefont {A.~K.}\ \bibnamefont
  {Biswas}}, \bibinfo {author} {\bibfnamefont {G.}~\bibnamefont {Compagno}},
  \bibinfo {author} {\bibfnamefont {G.~M.}\ \bibnamefont {Palma}}, \bibinfo
  {author} {\bibfnamefont {R.}~\bibnamefont {Passante}},\ and\ \bibinfo
  {author} {\bibfnamefont {F.}~\bibnamefont {Persico}},\ }\href
  {https://doi.org/10.1103/physreva.42.4291} {\bibfield  {journal} {\bibinfo
  {journal} {Physical Review A}\ }\textbf {\bibinfo {volume} {42}},\ \bibinfo
  {pages} {4291} (\bibinfo {year} {1990})}\BibitemShut {NoStop}%
\bibitem [{\citenamefont {Buchholz}\ and\ \citenamefont
  {Yngvason}(1994)}]{Buchholz1994}%
  \BibitemOpen
  \bibfield  {author} {\bibinfo {author} {\bibfnamefont {D.}~\bibnamefont
  {Buchholz}}\ and\ \bibinfo {author} {\bibfnamefont {J.}~\bibnamefont
  {Yngvason}},\ }\href {https://doi.org/10.1103/physrevlett.73.613} {\bibfield
  {journal} {\bibinfo  {journal} {Physical Review Letters}\ }\textbf {\bibinfo
  {volume} {73}},\ \bibinfo {pages} {613} (\bibinfo {year} {1994})}\BibitemShut
  {NoStop}%
\bibitem [{\citenamefont {Keller}(2000)}]{Keller2000}%
  \BibitemOpen
  \bibfield  {author} {\bibinfo {author} {\bibfnamefont {O.}~\bibnamefont
  {Keller}},\ }\href {https://doi.org/10.1103/physreva.62.022111} {\bibfield
  {journal} {\bibinfo  {journal} {Physical Review A}\ }\textbf {\bibinfo
  {volume} {62}},\ \bibinfo {pages} {022111} (\bibinfo {year}
  {2000})}\BibitemShut {NoStop}%
\bibitem [{\citenamefont {Chan}\ \emph {et~al.}(2002)\citenamefont {Chan},
  \citenamefont {Law},\ and\ \citenamefont {Eberly}}]{Chan2002}%
  \BibitemOpen
  \bibfield  {author} {\bibinfo {author} {\bibfnamefont {K.~W.}\ \bibnamefont
  {Chan}}, \bibinfo {author} {\bibfnamefont {C.~K.}\ \bibnamefont {Law}},\ and\
  \bibinfo {author} {\bibfnamefont {J.~H.}\ \bibnamefont {Eberly}},\ }\href
  {https://doi.org/10.1103/physrevlett.88.100402} {\bibfield  {journal}
  {\bibinfo  {journal} {Physical Review Letters}\ }\textbf {\bibinfo {volume}
  {88}},\ \bibinfo {pages} {100402} (\bibinfo {year} {2002})}\BibitemShut
  {NoStop}%
\bibitem [{\citenamefont {Fedorov}\ \emph {et~al.}(2005)\citenamefont
  {Fedorov}, \citenamefont {Efremov}, \citenamefont {Kazakov}, \citenamefont
  {Chan}, \citenamefont {Law},\ and\ \citenamefont {Eberly}}]{Fedorov2005}%
  \BibitemOpen
  \bibfield  {author} {\bibinfo {author} {\bibfnamefont {M.~V.}\ \bibnamefont
  {Fedorov}}, \bibinfo {author} {\bibfnamefont {M.~A.}\ \bibnamefont
  {Efremov}}, \bibinfo {author} {\bibfnamefont {A.~E.}\ \bibnamefont
  {Kazakov}}, \bibinfo {author} {\bibfnamefont {K.~W.}\ \bibnamefont {Chan}},
  \bibinfo {author} {\bibfnamefont {C.~K.}\ \bibnamefont {Law}},\ and\ \bibinfo
  {author} {\bibfnamefont {J.~H.}\ \bibnamefont {Eberly}},\ }\href
  {https://doi.org/10.1103/physreva.72.032110} {\bibfield  {journal} {\bibinfo
  {journal} {Physical Review A}\ }\textbf {\bibinfo {volume} {72}},\ \bibinfo
  {pages} {032110} (\bibinfo {year} {2005})}\BibitemShut {NoStop}%
\bibitem [{\citenamefont {Debierre}\ \emph
  {et~al.}(2015{\natexlab{a}})\citenamefont {Debierre}, \citenamefont {Durt},
  \citenamefont {Nicolet},\ and\ \citenamefont {Zolla}}]{Debierre2015a}%
  \BibitemOpen
  \bibfield  {author} {\bibinfo {author} {\bibfnamefont {V.}~\bibnamefont
  {Debierre}}, \bibinfo {author} {\bibfnamefont {T.}~\bibnamefont {Durt}},
  \bibinfo {author} {\bibfnamefont {A.}~\bibnamefont {Nicolet}},\ and\ \bibinfo
  {author} {\bibfnamefont {F.}~\bibnamefont {Zolla}},\ }\href
  {https://doi.org/10.1016/j.physleta.2015.06.011} {\bibfield  {journal}
  {\bibinfo  {journal} {Physics Letters A}\ }\textbf {\bibinfo {volume}
  {379}},\ \bibinfo {pages} {2577} (\bibinfo {year}
  {2015}{\natexlab{a}})}\BibitemShut {NoStop}%
\bibitem [{\citenamefont {Debierre}\ \emph
  {et~al.}(2015{\natexlab{b}})\citenamefont {Debierre}, \citenamefont
  {Goessens}, \citenamefont {Brainis},\ and\ \citenamefont
  {Durt}}]{Debierre2015}%
  \BibitemOpen
  \bibfield  {author} {\bibinfo {author} {\bibfnamefont {V.}~\bibnamefont
  {Debierre}}, \bibinfo {author} {\bibfnamefont {I.}~\bibnamefont {Goessens}},
  \bibinfo {author} {\bibfnamefont {E.}~\bibnamefont {Brainis}},\ and\ \bibinfo
  {author} {\bibfnamefont {T.}~\bibnamefont {Durt}},\ }\href
  {https://doi.org/10.1103/physreva.92.023825} {\bibfield  {journal} {\bibinfo
  {journal} {Physical Review A}\ }\textbf {\bibinfo {volume} {92}},\ \bibinfo
  {pages} {023825} (\bibinfo {year} {2015}{\natexlab{b}})}\BibitemShut
  {NoStop}%
\bibitem [{\citenamefont {Debierre}\ and\ \citenamefont
  {Durt}(2016)}]{Debierre2016}%
  \BibitemOpen
  \bibfield  {author} {\bibinfo {author} {\bibfnamefont {V.}~\bibnamefont
  {Debierre}}\ and\ \bibinfo {author} {\bibfnamefont {T.}~\bibnamefont
  {Durt}},\ }\href {https://doi.org/10.1103/physreva.93.023847} {\bibfield
  {journal} {\bibinfo  {journal} {Physical Review A}\ }\textbf {\bibinfo
  {volume} {93}},\ \bibinfo {pages} {023847} (\bibinfo {year}
  {2016})}\BibitemShut {NoStop}%
\bibitem [{\citenamefont {Debierre}\ \emph {et~al.}(2018)\citenamefont
  {Debierre}, \citenamefont {Stout},\ and\ \citenamefont
  {Durt}}]{Debierre2018}%
  \BibitemOpen
  \bibfield  {author} {\bibinfo {author} {\bibfnamefont {V.}~\bibnamefont
  {Debierre}}, \bibinfo {author} {\bibfnamefont {B.}~\bibnamefont {Stout}},\
  and\ \bibinfo {author} {\bibfnamefont {T.}~\bibnamefont {Durt}},\ }\href@noop
  {} {\bibfield  {journal} {\bibinfo  {journal} {arXiv:1809.11070}\ } (\bibinfo
  {year} {2018})}\BibitemShut {NoStop}%
\bibitem [{\citenamefont {Bia{\l}ynicki-Birula}(1996)}]{BialynickiBirula1996}%
  \BibitemOpen
  \bibfield  {author} {\bibinfo {author} {\bibfnamefont {I.}~\bibnamefont
  {Bia{\l}ynicki-Birula}},\ }in\ \href
  {https://doi.org/10.1016/s0079-6638(08)70316-0} {\emph {\bibinfo {booktitle}
  {Progress in Optics}}},\ \bibinfo {editor} {edited by\ \bibinfo {editor}
  {\bibfnamefont {E.}~\bibnamefont {Wolf}}}\ (\bibinfo  {publisher}
  {Elsevier},\ \bibinfo {year} {1996})\BibitemShut {NoStop}%
\bibitem [{\citenamefont {Federico}\ and\ \citenamefont
  {Jauslin}(2023{\natexlab{b}})}]{Federico2023}%
  \BibitemOpen
  \bibfield  {author} {\bibinfo {author} {\bibfnamefont {M.}~\bibnamefont
  {Federico}}\ and\ \bibinfo {author} {\bibfnamefont {H.~R.}\ \bibnamefont
  {Jauslin}},\ }\href {https://doi.org/10.1088/1751-8121/acd155} {\bibfield
  {journal} {\bibinfo  {journal} {Journal of Physics A: Mathematical and
  Theoretical}\ }\textbf {\bibinfo {volume} {56}},\ \bibinfo {pages} {235302}
  (\bibinfo {year} {2023}{\natexlab{b}})}\BibitemShut {NoStop}%
\bibitem [{\citenamefont {Compagno}\ \emph {et~al.}(1990)\citenamefont
  {Compagno}, \citenamefont {Passante},\ and\ \citenamefont
  {Persico}}]{Compagno1990}%
  \BibitemOpen
  \bibfield  {author} {\bibinfo {author} {\bibfnamefont {G.}~\bibnamefont
  {Compagno}}, \bibinfo {author} {\bibfnamefont {R.}~\bibnamefont {Passante}},\
  and\ \bibinfo {author} {\bibfnamefont {F.}~\bibnamefont {Persico}},\ }\href
  {https://doi.org/10.1080/09500349014551511} {\bibfield  {journal} {\bibinfo
  {journal} {Journal of Modern Optics}\ }\textbf {\bibinfo {volume} {37}},\
  \bibinfo {pages} {1377} (\bibinfo {year} {1990})}\BibitemShut {NoStop}%
\bibitem [{\citenamefont {Compagno}\ \emph {et~al.}(1995)\citenamefont
  {Compagno}, \citenamefont {Passante},\ and\ \citenamefont
  {Persico}}]{Compagno1995}%
  \BibitemOpen
  \bibfield  {author} {\bibinfo {author} {\bibfnamefont {G.}~\bibnamefont
  {Compagno}}, \bibinfo {author} {\bibfnamefont {R.}~\bibnamefont {Passante}},\
  and\ \bibinfo {author} {\bibfnamefont {F.}~\bibnamefont {Persico}},\
  }\href@noop {} {\emph {\bibinfo {title} {Atom-Field Interactions and Dressed
  Atoms}}}\ (\bibinfo  {publisher} {Cambridge University Press},\ \bibinfo
  {year} {1995})\BibitemShut {NoStop}%
\bibitem [{\citenamefont {Shirokov}(1978)}]{Shirokov1978}%
  \BibitemOpen
  \bibfield  {author} {\bibinfo {author} {\bibfnamefont {M.~I.}\ \bibnamefont
  {Shirokov}},\ }\href {https://doi.org/10.1070/pu1978v021n04abeh005541}
  {\bibfield  {journal} {\bibinfo  {journal} {Soviet Physics Uspekhi}\ }\textbf
  {\bibinfo {volume} {21}},\ \bibinfo {pages} {345} (\bibinfo {year}
  {1978})}\BibitemShut {NoStop}%
\bibitem [{\citenamefont {Landau}\ and\ \citenamefont
  {Peierls}(1930)}]{Landau1930}%
  \BibitemOpen
  \bibfield  {author} {\bibinfo {author} {\bibfnamefont {L.}~\bibnamefont
  {Landau}}\ and\ \bibinfo {author} {\bibfnamefont {R.}~\bibnamefont
  {Peierls}},\ }\href@noop {} {\bibfield  {journal} {\bibinfo  {journal}
  {Zeitschrift f{\"u}r Physik}\ }\textbf {\bibinfo {volume} {62}},\ \bibinfo
  {pages} {188} (\bibinfo {year} {1930})}\BibitemShut {NoStop}%
\bibitem [{\citenamefont {Mandel}\ and\ \citenamefont
  {Wolf}(1995)}]{Mandel1995}%
  \BibitemOpen
  \bibfield  {author} {\bibinfo {author} {\bibfnamefont {L.}~\bibnamefont
  {Mandel}}\ and\ \bibinfo {author} {\bibfnamefont {E.}~\bibnamefont {Wolf}},\
  }\href@noop {} {\emph {\bibinfo {title} {Optical Coherence and Quantum
  Optics}}}\ (\bibinfo  {publisher} {Cambridge University Press},\ \bibinfo
  {year} {1995})\BibitemShut {NoStop}%
\bibitem [{\citenamefont {Friedrichs}(1948)}]{Friedrichs1948}%
  \BibitemOpen
  \bibfield  {author} {\bibinfo {author} {\bibfnamefont {K.~O.}\ \bibnamefont
  {Friedrichs}},\ }\href {https://doi.org/10.1002/cpa.3160010404} {\bibfield
  {journal} {\bibinfo  {journal} {Communications on Pure and Applied
  Mathematics}\ }\textbf {\bibinfo {volume} {1}},\ \bibinfo {pages} {361}
  (\bibinfo {year} {1948})}\BibitemShut {NoStop}%
\bibitem [{\citenamefont {Lee}(1954)}]{Lee1954}%
  \BibitemOpen
  \bibfield  {author} {\bibinfo {author} {\bibfnamefont {T.~D.}\ \bibnamefont
  {Lee}},\ }\href {https://doi.org/10.1103/physrev.95.1329} {\bibfield
  {journal} {\bibinfo  {journal} {Physical Review}\ }\textbf {\bibinfo {volume}
  {95}},\ \bibinfo {pages} {1329} (\bibinfo {year} {1954})}\BibitemShut
  {NoStop}%
\bibitem [{\citenamefont {Jackson}(1999)}]{Jackson1999}%
  \BibitemOpen
  \bibfield  {author} {\bibinfo {author} {\bibfnamefont {J.~D.}\ \bibnamefont
  {Jackson}},\ }\href
  {https://www.ebook.de/de/product/3240907/jd_jackson_classical_electrodynamics_3e.html}
  {\emph {\bibinfo {title} {Classical Electrodynamics}}},\ \bibinfo {edition}
  {3rd}\ ed.\ (\bibinfo  {publisher} {John Wiley {\&} Sons Inc},\ \bibinfo
  {year} {1999})\BibitemShut {NoStop}%
\bibitem [{\citenamefont {Steck}(2023)}]{Steck2023}%
  \BibitemOpen
  \bibfield  {author} {\bibinfo {author} {\bibfnamefont {D.~A.}\ \bibnamefont
  {Steck}},\ }\href@noop {} {\emph {\bibinfo {title} {Quantum and Atom
  Optics}}}\ (\bibinfo  {publisher} {Available online at
  {\url{http://steck.us/teaching}}},\ \bibinfo {year} {2023})\ \bibinfo {note}
  {revision 0.13.21}\BibitemShut {NoStop}%
\bibitem [{\citenamefont {Licht}(1963)}]{Licht1963}%
  \BibitemOpen
  \bibfield  {author} {\bibinfo {author} {\bibfnamefont {A.~L.}\ \bibnamefont
  {Licht}},\ }\href {https://doi.org/10.1063/1.1703925} {\bibfield  {journal}
  {\bibinfo  {journal} {Journal of Mathematical Physics}\ }\textbf {\bibinfo
  {volume} {4}},\ \bibinfo {pages} {1443} (\bibinfo {year} {1963})}\BibitemShut
  {NoStop}%
\bibitem [{\citenamefont {Licht}(1966)}]{Licht1966}%
  \BibitemOpen
  \bibfield  {author} {\bibinfo {author} {\bibfnamefont {A.~L.}\ \bibnamefont
  {Licht}},\ }\href {https://doi.org/10.1063/1.1705079} {\bibfield  {journal}
  {\bibinfo  {journal} {Journal of Mathematical Physics}\ }\textbf {\bibinfo
  {volume} {7}},\ \bibinfo {pages} {1656} (\bibinfo {year} {1966})}\BibitemShut
  {NoStop}%
\bibitem [{\citenamefont {De~Bièvre}(2006)}]{DeBievre2006}%
  \BibitemOpen
  \bibfield  {author} {\bibinfo {author} {\bibfnamefont {S.}~\bibnamefont
  {De~Bièvre}},\ }in\ \href {https://doi.org/10.1007/3-540-32579-4_2} {\emph
  {\bibinfo {booktitle} {Large Coulomb Systems}}}\ (\bibinfo  {publisher}
  {Springer Berlin Heidelberg},\ \bibinfo {year} {2006})\ pp.\ \bibinfo {pages}
  {15--61}\BibitemShut {NoStop}%
\bibitem [{\citenamefont {{De Bi{\`{e}}vre}}(2007)}]{DeBievre2007}%
  \BibitemOpen
  \bibfield  {author} {\bibinfo {author} {\bibfnamefont {S.}~\bibnamefont {{De
  Bi{\`{e}}vre}}},\ }in\ \href {https://doi.org/10.1007/978-93-86279-33-0_5}
  {\emph {\bibinfo {booktitle} {Contributions in Mathematical Physics}}}\
  (\bibinfo  {publisher} {Hindustan Book Agency},\ \bibinfo {year} {2007})\
  pp.\ \bibinfo {pages} {123--146}\BibitemShut {NoStop}%
\bibitem [{\citenamefont {Federico}\ \emph {et~al.}(2022)\citenamefont
  {Federico}, \citenamefont {Dorier}, \citenamefont {Gu{\'{e}}rin},\ and\
  \citenamefont {Jauslin}}]{Federico2022}%
  \BibitemOpen
  \bibfield  {author} {\bibinfo {author} {\bibfnamefont {M.}~\bibnamefont
  {Federico}}, \bibinfo {author} {\bibfnamefont {V.}~\bibnamefont {Dorier}},
  \bibinfo {author} {\bibfnamefont {S.}~\bibnamefont {Gu{\'{e}}rin}},\ and\
  \bibinfo {author} {\bibfnamefont {H.~R.}\ \bibnamefont {Jauslin}},\ }\href
  {https://doi.org/10.1088/1361-6455/ac7e0e} {\bibfield  {journal} {\bibinfo
  {journal} {Journal of Physics B: Atomic, Molecular and Optical Physics}\
  }\textbf {\bibinfo {volume} {55}},\ \bibinfo {pages} {174002} (\bibinfo
  {year} {2022})}\BibitemShut {NoStop}%
\bibitem [{\citenamefont {Garrison}\ and\ \citenamefont
  {Chiao}(2008)}]{Garrison2008}%
  \BibitemOpen
  \bibfield  {author} {\bibinfo {author} {\bibfnamefont {J.~C.}\ \bibnamefont
  {Garrison}}\ and\ \bibinfo {author} {\bibfnamefont {R.~Y.}\ \bibnamefont
  {Chiao}},\ }\href@noop {} {\emph {\bibinfo {title} {Quantum Optics}}}\
  (\bibinfo  {publisher} {Oxford University Press},\ \bibinfo {year}
  {2008})\BibitemShut {NoStop}%
\bibitem [{\citenamefont {Fabre}\ and\ \citenamefont
  {Treps}(2020)}]{Fabre2020}%
  \BibitemOpen
  \bibfield  {author} {\bibinfo {author} {\bibfnamefont {C.}~\bibnamefont
  {Fabre}}\ and\ \bibinfo {author} {\bibfnamefont {N.}~\bibnamefont {Treps}},\
  }\href {https://doi.org/10.1103/RevModPhys.92.035005} {\bibfield  {journal}
  {\bibinfo  {journal} {Reviews of Modern Physics}\ }\textbf {\bibinfo {volume}
  {92}},\ \bibinfo {pages} {035005} (\bibinfo {year} {2020})}\BibitemShut
  {NoStop}%
\bibitem [{\citenamefont {Varshalovich}(1988)}]{Varshalovich1988}%
  \BibitemOpen
  \bibfield  {author} {\bibinfo {author} {\bibfnamefont {D.~A.}\ \bibnamefont
  {Varshalovich}},\ }\href@noop {} {\emph {\bibinfo {title} {Quantum Theory of
  Angular Momentum}}}\ (\bibinfo  {publisher} {World Scientific Pub.},\
  \bibinfo {year} {1988})\BibitemShut {NoStop}%
\bibitem [{\citenamefont {Dai}\ \emph {et~al.}(2012)\citenamefont {Dai},
  \citenamefont {Kamionkowski},\ and\ \citenamefont {Jeong}}]{Dai2012}%
  \BibitemOpen
  \bibfield  {author} {\bibinfo {author} {\bibfnamefont {L.}~\bibnamefont
  {Dai}}, \bibinfo {author} {\bibfnamefont {M.}~\bibnamefont {Kamionkowski}},\
  and\ \bibinfo {author} {\bibfnamefont {D.}~\bibnamefont {Jeong}},\ }\href
  {https://doi.org/10.1103/physrevd.86.125013} {\bibfield  {journal} {\bibinfo
  {journal} {Physical Review D}\ }\textbf {\bibinfo {volume} {86}},\ \bibinfo
  {pages} {125013} (\bibinfo {year} {2012})}\BibitemShut {NoStop}%
\bibitem [{\citenamefont {Appel}(2007)}]{Appel2007}%
  \BibitemOpen
  \bibfield  {author} {\bibinfo {author} {\bibfnamefont {W.}~\bibnamefont
  {Appel}},\ }\href@noop {} {\emph {\bibinfo {title} {Mathematics for Physics
  and Physicists}}}\ (\bibinfo  {publisher} {Princeton University Press},\
  \bibinfo {year} {2007})\ p.\ \bibinfo {pages} {672}\BibitemShut {NoStop}%
\end{thebibliography}%

\end{document}